\documentclass[]{zakoproc}
\usepackage{graphicx}
\usepackage{subfigure}
\begin{document}

\title{Probing the Magnetic Fields of Nearby Spiral Galaxies at Low Frequencies with LOFAR}
\author{David D. Mulcahy$^{1}$  Robert Drzazga$^{2}$\\ Bj\"orn Adebahr$^{3}$ James Anderson$^{1}$
Rainer Beck$^{1}$\\ Michael R. Bell$^{4}$ Krzysztof Chyzy$^{2}$ Ren\'e Gie\ss\"ubel$^{1}$ \\ George Heald$^{5}$ Andreas Horneffer$^{1}$ Wojciech Jurusik$^{2}$ \\  Roberto Pizzo$^{5}$
Anna Scaife$^{6}$ Carlos Sotomayor-Beltran$^{3}$ \\ Blazej Nikiel-Wroczynski$^{2}$\\
on behalf of the LOFAR Magnetism Key Science Project \& \\ the LOFAR collaboration}

\institute{$^{1}$Max-Planck-Institut f\"ur Radioastronomie, Auf dem H\"ugel 69, 53121 Bonn, Germany\\
           $^{2}$Astronomical Observatory, Jagiellonian University, ul. Orla 171, 30-244 Krak\'ow, Poland\\
           $^{3}$Ruhr-Universit\"at Bochum, Universit\"atsstrasse 150, 44801 Bochum, Germany\\
           $^{4}$Max-Planck-Institut f\"ur Astrophysik, Karl-Schwarzschild-Str. 1, 85748 Garching, Germany\\
           $^{5}$ASTRON, Postbus 2, 7990 AA Dwingeloo, The Netherlands\\
           $^{6}$School of Physics \& Astronomy, University of Southampton, Southampton, Hampshire, SO17 1BJ, United Kingdom\\}
\markboth{D.Mulcahy, R.Drzazga et al.}{Probing Magnetic Fields of Nearby 
Spiral Galaxies at Low Frequencies with LOFAR \ldots}

\maketitle

\begin{abstract}

While the Low Frequency Array (LOFAR) is still in its commissioning
phase, early science results are starting to emerge. Two nearby
galaxies, M51 and NGC4631, have been observed as part of the
Magnetism Key Science Project's (MKSP) effort to increase our
understanding of the nature of weak magnetic fields in galaxies.
LOFAR and the complexity of its calibration as well as the aims \&
goals of the MKSP are presented.

\end{abstract}

\section{Introduction to the Low Frequency Array}

LOFAR is a next generation radio telescope
currently in its construction \& commissioning phase. Its core is
located in the Netherlands and the international stations are located in Germany,
France, Sweden and the UK, with plans to expand to additional
European countries. The LOFAR telescope
consists of two distinct types of antennas, namely the Low Band
Antenna (LBA) which operate between 10 and 90\,MHz and the High
Band Antenna (HBA) which operates between 120\,MHz and 240\,MHz. A
single LBA antenna is able to see the entire sky and the telescope is pointed by
forming beams at each station. Single HBA tiles (each of which contain 16 dipole pairs)
have a large field of view but are not able to observe the entire sky (Stappers et al. 2011 \& van Haarlem, Wise, et al. in prep).
The data from all stations are then correlated at a central supercomputer.

LOFAR, due to its bandwidth and low frequency, can
measure Faraday rotation measures (RM) with a very high precision.
We will use ``RM Synthesis'' (Brentjens \& de Bruyn 2005) and thereby
detect weak magnetic fields and low electron densities in galaxies
which are unobservable at higher frequencies. The high RM precision
of LOFAR will also help to study RMs from pulsars, stellar and AGN
jets, and the lobes of radio galaxies with unprecedented accuracy.

\section{Introduction to the MKSP}

The main aim of the international LOFAR Key Science Project on
Cosmic Magnetism in the Nearby Universe (MKSP) is to observe the
magnetic fields in the Milky Way and nearby galaxies by observing
total and polarized radio synchrotron emission at very low
frequencies. Low-frequency radio emission traces low-energy
cosmic-ray electrons which suffer less from energy losses and can
propagate further away from their acceleration sites into regions with weak
magnetic fields. Trying to detect polarization at these low frequencies
is a challenge in itself and can be best done through RM Synthesis.
RM Synthesis will allow the investigation of the 3-D structure of
magnetic fields in the Milky Way and in nearby galaxies. We
plan to use LOFAR to detect the radio emission in galaxies at
distances of tens of kpc from the star-forming disks and study structures like extended
gaseous halos of spiral and dwarf irregulars, or tails of
interacting and stripped spirals. The MKSP encompasses several research 
groups focused on: the magnetic field of the Milky Way through continuum observations as
well as pulsar Rotation Measures (RM); the magnetic fields of nearby
galaxies and giant radio galaxies; intergalactic filaments \& finally stellar jets.

RM Synthesis has already been used on LOFAR data to successfully detect
polarization. Emission in Stokes Q \& U was seen
alternating for the pulsar PSRJ0218+4232 (Heald et al. 2011). RM Synthesis was performed
and showed that the pulsar has a Faraday Depth of about
61\,rad/m$^{2}$ which agrees with published data. Currently, efforts
are concentrating on finding pulsars that could be used as
polarized calibrators.

\section{The calibration challenges of LOFAR data}

The calibration and imaging of LOFAR data is a great challenge, in part because at these
frequencies there can be large ionospheric phase variations which
distort the observed brightness distribution in uncorrected images.
An overview of the imaging pipeline can be seen in Heald et al. (2011).
The ionospheric phase variations can be corrected for by iterative self calibration using bright radio sources in the field.
Strong sources nearby can distort the target source. These problematic sources, however, can
be subtracted from the uv data.

The LOFAR antennas detect the emission from bright sources even when located far away from the station beam pointing direction.
This can distort the images of the target field. 
Most problematic are the very bright ``A-Team'' sources such as Cygnus A and Cassiopeia A. 
Fortunately, there is now software in place to subtract their
distorting emission through a process called ``demixing'' (van der Tol, S. et al. 2007), 
which is done prior to calibration and is essential for all LBA observations.

Calibration with LOFAR is done through the program BBS which
requires a Local Sky Model (LSM) which is
extracted from a Global Sky Model (GSM) that is stored in the
database. The initial GSM development will come about through the
Multifrequency Snapshot Sky Survey (MSSS) (MSSS; Heald et al. in prep) 
which has started in October 2011. At present,
skymodels are created from the WENSS or VLSS surveys.
BBS calibrates the complex station gains using this LSM.

LOFAR has the novel ability in that it is able to observe two or more parts of
the sky simultaneously. This has the advantage of continuously
observing the target source and getting the maximum uv coverage
whilst observing a calibrator. The total bandwidth is
divided up by the number of sources observed, in these cases the
on-source bandwidth is halved. The calibrator can be then easily calibrated
and the obtained gain and phase solutions found can be transferred to
the target observation.

\section{Dual observations of 3C295 \& M51}

The spiral galaxy M51 was chosen to be observed by the MKSP due to
the high polarization degree in the interarm regions with structures
as large as 5\,kpc in size, as seen in Fletcher et al. (2011). 3C295
and the relatively face-on spiral galaxy M51 were observed
simultaneously for a 5 hour period in May 2011 during nighttime to
minimize ionospheric distortions.

Figure~1, left image, shows M51 at 145.12\,MHz calibrated by using the transfer
of gain solutions method described in section 3. The rms for this image is approximately 7mJy/beam which can be improved for a single subband by the subtraction of nearby sources.
As one can see from comparing the radio image to the optical image, an extended disk is very evident as well
as the inner western spiral arm. The image also looks very similar to an image obtained by Segalovitz (1976) at 610\,MHz.
However greater resolution is achieved in the LOFAR image with a beamsize of 61$\times$30 arcsecs. Robust weighting of 0.25 was used.

\begin{figure}[h!]
\begin{center}$
\begin{array}{cc}
\includegraphics[scale=.3]{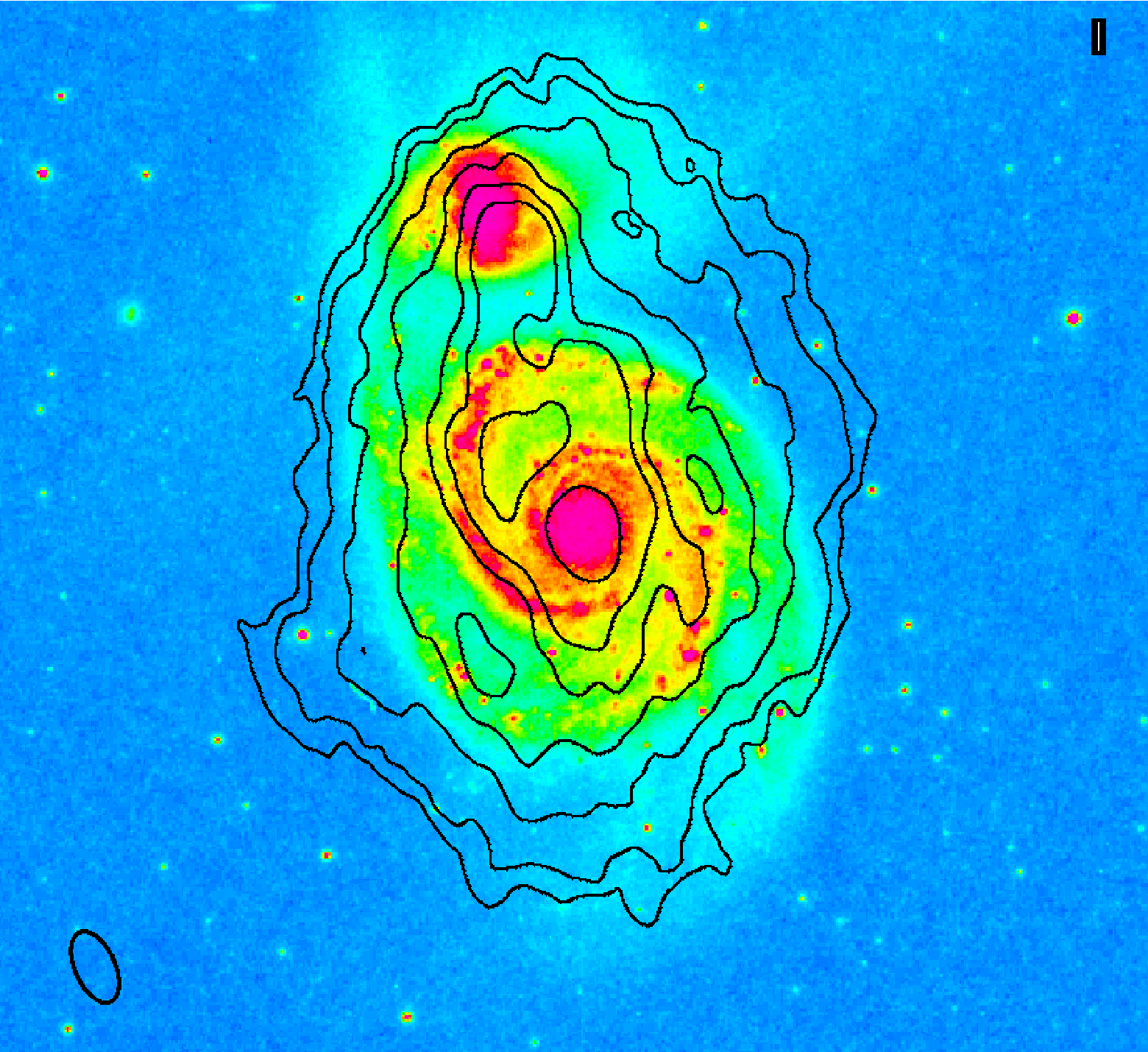} &
\includegraphics[scale=.5]{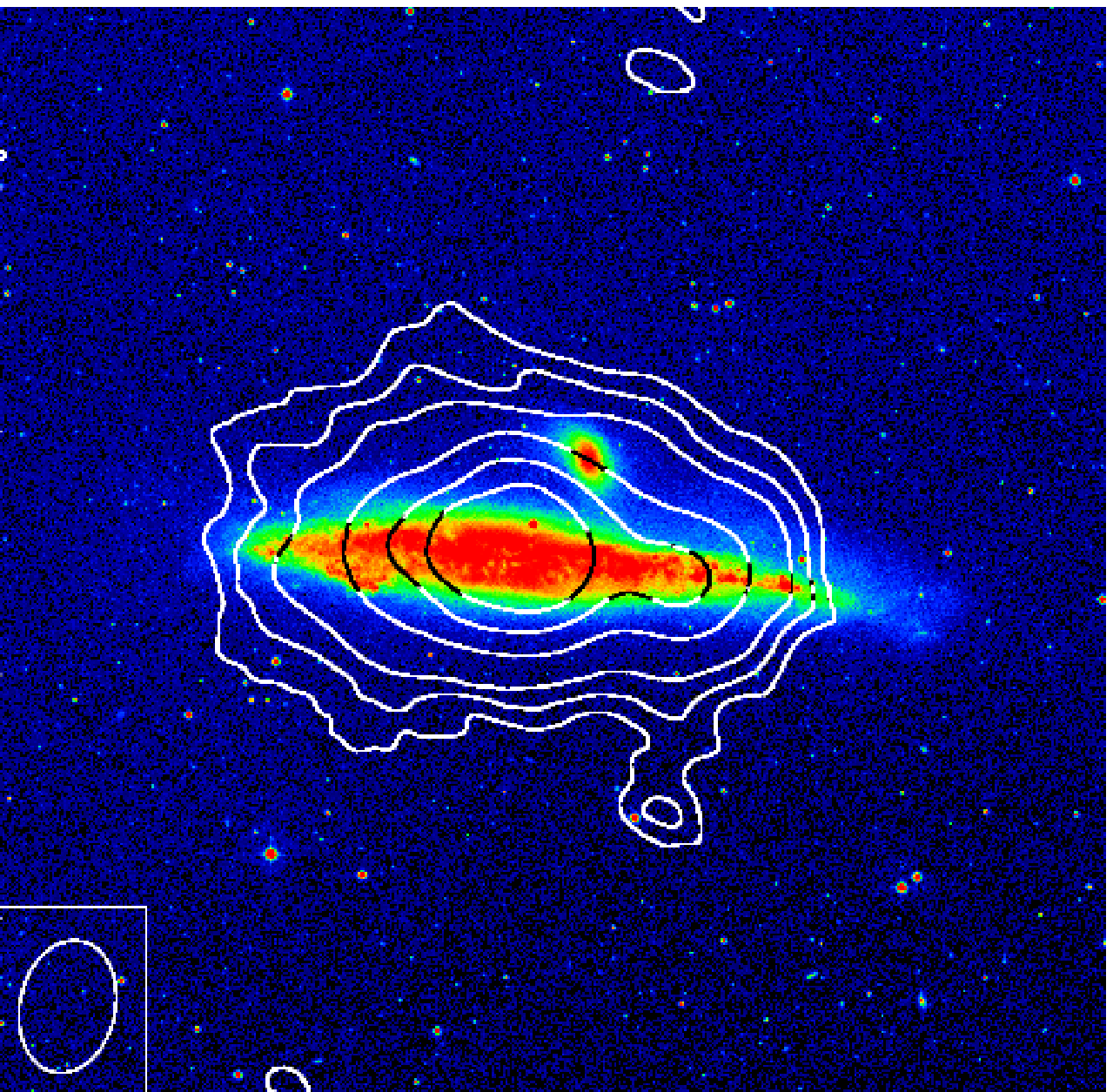}
\end{array}$
\end{center}
\caption{Left: M51 for a single subband at 145.12\,MHz (contours) with a bandwidth of approximately 400\,KHz overlayed onto an optical DSS image. (D.Mulcahy).
Right: NGC4631 for a single subband at 157\,MHz (bandwidth is approximately 400\,KHz) (contours) calibrated from a single source in the
field, overlayed onto an optical blue DSS image (R.Drzazga).}
\end{figure}

\section{Dual observations of 3C286 \& NGC4631}

3C286 and the edge-on galaxy NGC4631 were also simultaneously observed in May 2011
during nighttime. The purpose of observing this galaxy was to detect the
extended radio halo known to exist seen in Golla \& Hummel (1994).
The large thickness of the radio halo is due to the high star formation efficiency rate 
in the disk as well as gravitational interaction with NGC4656 which can be seen from Krause (this volume). 
Figure~1, right image, shows an image of NGC4631 that has been calibrated with BBS
on a single source in the field itself (4C+32.40) with a total
flux of 3.2\,Jy (at 157\,MHz), interpolated between NVSS and VLSS.
This was then followed by two selfcal cycles. The rms noise is 20mJy/beam which again can be improved within a subband. The beam is 165$\times$118 arcsecs.
A very extended synchrotron halo can be seen which is comparable to the image from Hummel \& Dettmar (1990) at 327\,MHz.
Efforts are currently being made to create images using the transfer of gains solutions method.

\section{Conclusions}

From these two observations, it is clear that LOFAR has started to
give reliable images of extended sources. The
extended synchrotron disk and halo of M51 and NGC4631 are easily seen in our
images which were expected from observations in other low wavelengths seen in for M51 at 610\,MHz and for NGC4631 at 327\,MHz.
It is important to note that the presented images are from a single subband.
When the whole bandwidth of 24\,MHz (122 subbands) is combined, a significant increase in sensitivity is expected.
This should reveal the very outer regions of these galaxies.

A search for polarization with RM Synthesis is planned in the coming
months.

\acknowledgements{This research was performed in the framework of
the DFG Forschergruppe 1254 ``Magnetisation of Interstellar and
Intergalactic Media: The Prospects of Low-Frequency Radio
Observations''. -- LOFAR, the Low Frequency Array designed and
constructed by ASTRON, has facilities in several countries, that are
owned by various parties (each with their own funding sources), and
that are collectively operated by the International LOFAR Telescope
(ILT) foundation under a joint scientific policy.}

\end{document}